\documentclass[11pt, twoside, eqno]{article}
\usepackage{amssymb}
\usepackage{mathrsfs}
\usepackage{amsmath}
\usepackage{amsfonts}
\usepackage{amsthm}
\usepackage{subfigure}
\usepackage{graphicx}
\usepackage{booktabs}
\usepackage{setspace}

\allowdisplaybreaks \numberwithin{equation}{section}

\def\hs{\hspace{0.5cm}}

\newtheorem{cor}{\noindent\rm\bf Corollary}[section]
\newtheorem{thm}{\noindent\rm\bf Theorem}[section]

\newtheorem{prop}{\noindent\rm\bf Property}[section]
\begin{document}
%\doublespacing
\baselineskip=15pt
\renewcommand{\arraystretch}{2}
\arraycolsep=1pt
\title{\bf \Large Integral Cryptanalysis of the Block Cipher E2
\footnotetext{\hspace{-0.6cm} ${ }^{*}$ Corresponding authors. \\
   E-mail addresses: nlwt8988@gmail.com. }
\author{\vspace{-0.1cm}\bf Wentan Yi$^{*}$ and  Shaozhen Chen \\
\vspace{-0.5cm}\small\it State Key Laboratory of Mathematical Engineering and Advanced Computing,\\
\small\it Zhengzhou 450001, China }}
\date{}

\maketitle

\begin{center}
\begin{minipage}{15.2cm}
\small{\bf Abstract.}  Block cipher E2, designed and submitted by Nippon
Telegraph and Telephone Corporation, is a first-round Advanced Encryption
Standard candidate. It employs a Feistel structure as global structure and
two-layer substitution-permutation network structure in round function with initial
transformation $IT$ function before the first round and final transformation $FT$
function after the last round. The design principles influences several more recent block
ciphers including Camellia, an ISO/IEC standard cipher.

\quad In this paper, we focus on the key-recovery attacks on reduced-round E2-128/192 taking
both $IT$ and $FT$ functions in consideration with integral cryptanalysis. We first improve the
relations between zero-correlation linear approximations
and integral distinguishers, and then deduce some integral distinguishers from  zero-correlation
linear approximations over 6 rounds of E2. Furthermore,  we apply these integral distinguishers to break
6-round E2-128 with $2^{120}$ chosen plaintexts (CPs), $2^{115.4}$ encryptions and $2^{28}$ bytes memory.
In addition, the attack on 7-round E2-192 requires $2^{120}$ CPs, $2^{172.5}$ encryptions and $2^{60}$ bytes memory.
\medskip

\end{minipage}
\end{center}

\section{\large\bf Introduction}

Integral attack, extended from square attack [1], is one of the most popular cryptanalytic
tools for block ciphers. At FSE 2002, Knudsen and Wagner [2] introduced the
definition of integral attack and unified a kind of Square attack such as saturation
attack [3] and multiset attack [4] as integral attack. The basic idea of integral attack is to
analyze some properties of sums of values, such as zero-sum property in specific parts of
ciphertexts. Thus, it can be seen as a dual to differential cryptanalysis  [5]. However,
integral attack has not been thought suitable for bit-based block ciphers£¬ until Z¡¯aba et al. [6] proposed the bit-based integral
attack in 2008, which was applied to Noekeon, Serpent and PRESENT reduced up to 5, 6
and 7 rounds. Integral attack has been applied to many block ciphers so far, such as
Rijndael [7], ARIA [8] and MISTY1[9]  and it is also one of the best
attacks on AES [10]. It reveals that integral attacks may be an important tool  to
understand the security of block ciphers.

Linear cryptanalysis [11] is another prominent cryptanalysis method against block ciphers.
Several extensions of linear cryptanalysis have been introduced so far, such as multiple
linear approximations cryptanalysis [12] and multidimensional linear cryptanalysis [13]. In
2012, Bogdanov et al. [14] developed a new method for cryptanalysis of block  ciphers
named zero-correlation linear attack. Zero-correlation linear attack uses linear
approximations with zero-correlation for all keys. It can be seen as the
counterpart of impossible differential cryptanalysis[15].

A number of relations have been established among some previously known
statistical attacks on block ciphers so far.  Chabaud and Vaudenay  [16] presented the
mathematical links between differential probability and linear correlation. Relations
between multidimensional linear and truncated differential distinguishers were
established by Blondeau and Nyberg [17], and they showed that the existence of
zero-correlation relations is equivalent to the existence of an impossible differential property.
Integral attacks also have some relations with other statistical attacks. Integral and zero-correlation distinguishers were
established by Bogdanov et al. [18] and they presented that an integral implies a
zero-correlation distinguisher and a zero-correlation distinguisher implies an integral if
input and output masks are independent of each other.

E2 [19] is a 128-bit block cipher with a user key length of 128, 192 or 256 bits. For simplicity, we denote by
E2-128/192/256 the three versions of E2 that use 128, 192, and 256 key bits, respectively. It was designed
and submitted to Advanced Encryption Standard project by Nippon Telegraph and Telephone Corporation.
The design criteria of E2 are conservative, adopting a Feistel network structure as a global
structure and the two-layer Substitution-Permutation Network (SPN) structure in its round function.
All operations used in the data randomization phase are byte table lookups and byte xor¡¯s except
32-bit multiplications in initial transformation $IT$ and final transformation $FT$, which successfully
makes E2 a fast software cipher independent of target platforms.

The cryptanalytic results for round-reduced E2 have been concentrating
around truncated, impossible differentials and zero-correlation linear attack.
7-round truncated differential characteristic of E2 was proposed by Matsui and Tokita in [20],
 and then they proposed a possible attack on an 8-round E2 without $IT$ and $FT$ functions under data complexity
$2^{100}$ chosen plaintexts and unknown time complexity. Moriai et al.[21] found another 7-round
truncated differential characteristic with higher probability and they presented a possible key recovery
attack on 8-round E2-128 without $IT$ and $FT$  under data complexity $2^{94}$ chosen plaintexts and uncertain time complexity.

For the security  against impossible differential cryptanalysis,  Aoki et al.[22] studied the impossible
differentials of E2 for the first time. However, they found no impossible differential more than 5 rounds
for E2 without $IT$ and $FT$ functions. The authors[23] declared some 6-round impossible differential characteristics of E2 without $IT$ and $FT$ functions, but still no attack results on E2 was given.  Wei et al.[24] presented key recovery
attack on 7-round E2-128 without $IT$ and $FT$ requiring $2^{120}$ chosen plaintexts
and $2^{115.5}$ encryptions, and the key recovery attack on 8-round E2-256 without
$IT$ and $FT$ requires $2^{214}$ encryptions with $2^{121}$ chosen plaintexts. Recently, Wen et al [25]. identified zero-correlation linear approximations
over 6 rounds of E2, and then they introduced the multidimensional zero-correlation linear attacks on  8-round E2-128 and 9-round E2-256 without $IT$ and $FT$. In addition, they  proposed  key recovery attacks on 6-round E2-128 and 7-round E2-256 with both $IT$ and $FT$ taken into consideration for the first time. Attack results on E2 are summarized in Table 1.

\begin{table}[tbp]
\centering
\scriptsize
\begin{tabular}{ccccccc}
\hline
Attack Type &key size& Rounds & $IT/FT$& Date & Time & Source\\
\hline
\vspace{-0.1in}Impossible Differential & 128 & 7 & None &$2^{120}$CPs & $2^{115.5}$Enc & [24] \\
\vspace{-0.1in}multidimensional zero-correlation  & 128 & 8 & None &$2^{124.1}$CPs & $2^{119.1}$ Enc & [25] \\
\vspace{-0.1in}Truncated Differential & 128 & 8 & One &$2^{94}$CPs & $-$ & [21] \\
\vspace{-0.1in}Multidimensional zero-correlation  & 128 & 6 & Both&$2^{123.7}$KPs & $2^{119.1}$ Enc & [25] \\
Integral  & 128 & 6 & Both&$2^{120}$CPs & $2^{115.4}$ Enc & Sect.4\\
\hline
Integral  & 192 & 7 & Both&$2^{120}$CPs & $2^{172.5}$ Enc & Sect.4\\
\hline
\vspace{-0.1in}Impossible Differential & 256 & 8 & None &$2^{121}$CPs & $2^{214}$Enc & [24] \\
\vspace{-0.1in}Multidimensional Zero-Correlation &256& 9 & None &$2^{124.6}$ KPs &$2^{255.5}$Enc & [25]\\
\vspace{-0.1in}Multidimensional Zero-Correlation & 256 &7&Both &$ 2^{124.7}$ KPs &$2^{252.8}$Enc & [25]\\
Integral  & 256 & 7 & Both&$2^{120}$CPs & $2^{172.5}$ Enc & Sect.4\\
\hline
\end{tabular}

%KP refer to the number of  known plaintexts,\\
% Enc refers to the number of encryptions.
\caption{Summary of the attacks on E2}
\end{table}
In this paper, the 6-round integral distinguishers are discussed in detail. Furthermore, we  investigate the security
of reduced-round E2-128/192 with both $IT$ and $FT$ functions against integral cryptanalysis. Our contributions can be summarized as follows.

1. We  deduce some 6-round integral distinguishers from new-revealed 6-round zero-correlation linear approximations of E2 by improving the relations between zero-correlation linear approximations and integral distinguishers. The integral distinguishers have much stronger ability to distinguish the right from wrong keys.

2.  Integrals attacks on 6-round E2-128 and 7-round E2-192 with both $IT$ and $FT$ taken in consideration are proposed. To  my knowledge, they are the first integral attacks on reduced-round of E2 with both $IT$ and $FT$.

The paper is organized as follows. Section 2 gives a brief description of block cipher E2 and outlines the ideas of  zero-correlation linear cryptanalysis and integral cryptanalysis. In addition,  the relations between zero-correlation linear approximations and
integral distinguishers are also discussed. Some new zero-correlation linear approximations and the deduced integral distinguishers are shown in Section 3. Section 4 illustrate our attacks on 6-round E2-128 and the 7-round E2-192 with both $IT$ and $FT$ functions taken in consideration. We conclude in Section 5.

\section{\large \bf  Preliminarise}
\subsection{\bf Description of E2}
E2 is a 128-bit block cipher proposed by NTT in 1998 and is selected as one
of the fifteen candidates in the first round of AES project. E2 is a Feistel cipher
with two-layer SPN structure in its round function and iterates  12 times  with an $IT$ function
at the beginning and an $FT$ function at the end. Figure 1 (a) shows the outline of the E2 encryption
process, also see Algorithm 1. The decryption process of E2 is the same as the encryption process except for the order of the subkeys.
\begin{figure}
  \centering
  \includegraphics[width=12cm]{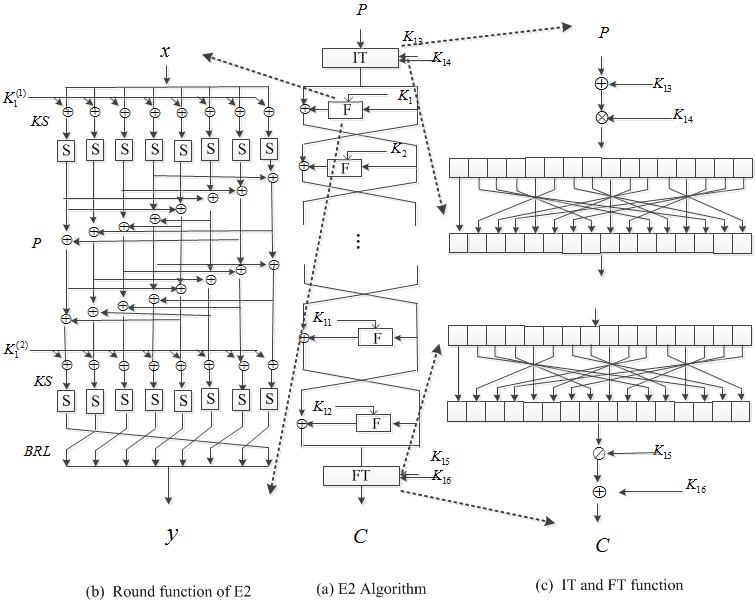}
  \caption{The structure and building blocks of E2}
\end{figure}

The round function employs SPS structures including the XOR operation with the first
round subkey, the first nonlinear transformation consisting of eight parallel $8\times 8$
S-boxes, the linear transformation $P$, the XOR operation with the second round
subkey, the second nonlinear transformation consisting of eight parallel $8 \times 8$ S-boxes and the second linear layer $BRL$ in order.  Figure 1(b) outlines the round function that consists of $S$-function, $P$-function, and $BRL$-function.
We refer [19] for more details of the specification and notations.

The first linear transformation $P : F^{64}_2 \rightarrow F^{64}_2$ in round function can be expressed with
matrix-vector product,

\begin{equation*}
\footnotesize
\left(\vspace{-0.2in}
\begin{array}{c}
 \vspace{-0.18in}z'_1 \\
 \vspace{-0.18in}z'_2 \\
 \vspace{-0.18in}z'_3\\
 \vspace{-0.18in}z'_4 \\
 \vspace{-0.18in}z'_5 \\
 \vspace{-0.18in}z'_6\\
 \vspace{-0.18in}z'_7 \\
  z'_8
\end{array}
\right) =\left(
\begin{array}{cccccccc}
 \vspace{-0.18in}0 & 1 & 1& 1 & 1 & 1 & 1 & 0 \\
 \vspace{-0.18in}1 & 0 & 1& 1 & 0 & 1 & 1 & 1 \\
 \vspace{-0.18in}1 & 1 & 0& 1 & 1 & 0 & 1 & 1 \\
 \vspace{-0.18in}1 & 1 & 1& 0 & 1 & 1 & 0 & 1 \\
 \vspace{-0.18in}1 & 1 & 0& 1 & 1 & 1 & 0 & 0 \\
 \vspace{-0.18in}1 & 1 & 1& 0 & 0 & 1 & 1 & 0 \\
 \vspace{-0.18in}0 & 1 & 1& 1 & 0 & 0 & 1 & 1 \\
1 & 0 & 1& 1 & 1 & 0 & 0 & 1 \\
\end{array}
\right)
\cdot
\left(
\begin{array}{ccc}
 \vspace{-0.18in}z_1 \\
 \vspace{-0.18in}z_2 \\
 \vspace{-0.18in}z_3\\
 \vspace{-0.18in}z_4 \\
 \vspace{-0.18in}z_5 \\
 \vspace{-0.18in}z_6\\
 \vspace{-0.18in}z_7 \\
  z_8
\end{array}
\right)
\end{equation*}
where $z_i, z'_i \in F_2^8$ and  $1\leq i \leq 8$.

The second linear layer $BRL$ is much easier and can be  represented as follows:
$$ BRL : F_{2^8}^8 \rightarrow F_{2^8}^8: (y_1, y_2,..., y_8)\rightarrow(y_2, y_3,..., y_8, y_1).$$

Both $IT$ and $FT$ functions are key-dependent transformations, which are designed to increase security
against both known attacks and unknown attacks. $IT$ and $FT$ are described as follows.
$$IT : F_2^{128}\times F_2^{128} \times F_2^{128}\rightarrow F_2^{128} ; (X, A, B)= BP((X \oplus A)\otimes B);$$
$$FT : F_2^{128}\times F_2^{128} \times F_2^{128}\rightarrow F_2^{128} ; (X, A, B)= (BP^{-1}(X)\oslash B)\oplus A,$$
where the byte permutation $BP$ and $BP^{-1}$ are shown in Figure 1(c). As to operation $\otimes$ and $\oslash$,
if we represent $X = (x_1, x_2, x_3, x_4), Y = (y_1, y_2, y_3, y_4)$,
and $B = (b_1, b_2, b_3, b_4)$, where $x_i, y_i, b_i \in F_2^{32}$; $1\leq i \leq 4$, and use $\vee 1$ to denote
bitwise logical OR with $1 \in F^{32}_2$ , then we have:
$$Y = X\otimes B := y_i = x_i(b_i\vee1) mod \,2^{32}\, (i = 1, 2, 3, 4);$$
\vspace{-0.25in}$$Y = X\oslash B := x_i = y_i(b_i\vee 1)^{-1} mod \,2^{32} \,(i = 1,2, 3, 4).$$
In addition, our attacks do not utilize the key relation, we omit the details of E2's key schedule.
\begin{table}[tbp]
\centering
\scriptsize
\begin{tabular}{l}
\hline
Algorithm 1 The E2 block cipher \\
\hline
\vspace{-0.1in}Require: 128-bit plaintext $P=(P_L, P_R)$; main key $K$,\\
\vspace{-0.1in}Ensure: 128-bit ciphertext $C=(C_L,C_R)$.\\
\vspace{-0.1in}1: Derive  round keys $K_i$, $(1 \leq i \leq 16)$ from $K$.\\
\vspace{-0.1in}2: $(L_0,R_0)=IT(P,K_{13},K_{14})$.\\
\vspace{-0.1in}3: for $j=1$ to 12 do\\
\vspace{-0.1in}4: $R_j = F(R_{j-1},K_j)\oplus L_{j-1}, L_j = R_{j-1},$ \\
\vspace{-0.1in}5: end for \\
\vspace{-0.1in}6: $(C_L,C_R)=FT((R_{12},L_{12}),K_{15},K_{16})$.\\
7: return $C=(C_L, C_R)$.\\
\hline
\end{tabular}
\end{table}
\subsection{\bf Basic ideas of integral and zero-correlation linear attack}

In this section, we briefly recall the basic concepts of integral cryptanalysis. Let $E = E_1 \cdot E_0$ be the encryption function of an $r$ round block cipher, where
$E_0$ is the first $r_1$ rounds of $E$ and $E_1$ is the last $r-r_1$ rounds. It can be written
formally as
$$C = E(P,K)= E_1(E_0(P,K_0),K_1),$$
or equivalently,
$$E_1^{-1}(C,K_1)= E_0(P,K_0),$$
where $E^{-1}_1$ is the inverse function of $E_1$, $K$ is the master key, $K_0$ and $K_1$ are
subkeys in the first $r_1$ rounds and the last $r - r_1$ rounds, respectively.

In integral attacks, an attacker first selects a set of ${2^d}$ plaintexts, where $d$ bit
positions of $P$ take on all values through the set and the other bits of $P$ are
chosen to be arbitrary constants. Then, some properties, such as a zero-sum property, of the set of plaintexts propagating through $r_1$ round encryptions is proved. For example, an attacker
demonstrates that
$$\sum_{p\in\Omega}E_0(p,K_0) mod \,2 =0,$$
where $\Omega$ is the set of $2^d$ plaintexts. Finally, the subkey $K_1$ in the last $r-r_1$ rounds
is guessed and equation

$$\sum_{p\in \Omega,c =E(p,K)}E^{-1}_1(c,K_1) mod \, 2 =0$$
is used to verify the guess. The remaining key bits in the master key $K$ will be
obtained by exhausting method.

Notice that, the integral distinguisher is built upon a specific parts
of the output of $E_0$, and the property of the output of $E_0$ determine the
ability of the distinguisher to sieve keys. In this paper, we  develop a stronger integral
distinguisher from some zero-correlation linear approximates.  We recall some basic concepts
of zero-correlation linear cryptanalysis.

Let $a$ and $\beta$ be the input and output masks. We denote $a\rightarrow \beta$ the correlation of a linear approximation of a vectorial function $f(x)$ by
$$\text{Cor}_{x}(\beta \cdot f(x), a\cdot x)=2\text{Pr}_{x}(\beta \cdot f(x)\oplus a\cdot x=0)-1,$$
where the scalar product of binary vectors is denoted by $a\cdot x = \oplus_{i=1}^{n}a_i x_i$. In zero-correlation linear cryptanalysis, the distinguisher uses linear approximations with zero correlation for
all keys, while the classical linear cryptanalysis utilizes linear approximations with correlation as far from zero as
possible. Let us recall a  result of correlations of restrictions of Boolean functions.
\begin{thm}
Let $f : F^{n_1}_2 \times F^{n_2}_2 \rightarrow F^{n}_2$ be a vectorial Boolean function, and let
$x_{n_1} \in F^{n_1}_2$ be uniformly distributed. Then
$$\sum_{x_{n_1}\in F^{n_1}_2}\text{cor}^2_{x_{n_2}}(a\cdot x_{n_2} \oplus b\cdot f(x_{n_1}, x_{n_2})) = 2^{n_2}\sum_{c\in F^{n_1}_2}
\text{cor}^2_{x_{n_1},x_{n_2}}( c  \cdot x_{n_1} \oplus a\cdot x_{n_2} \oplus b\cdot  f(x_{n_1}, x_{n_2})),$$
for all $a \in F^{n_2}_2$, $b \in F^{n}_2\setminus \{0\}$, where $n_1, n_2, n$ are positive integrals.
\end{thm}
This fact was named as Fundamental Theorem in [26], which describes
the links between integral distinguishers and the zero-correlation linear approximates.
In order to apply Theorem 2.1 to the analysis of E2 block cipher, we should make some
improvements. The following theorem can be proved by repeating the proof of Theorem 2.1 in [26].

\begin{thm}
Let $f : F^{m_1}_2 \times F^{m_2}_2 \times F^{m_3}_2 \rightarrow F^m_2$ be a vectorial Boolean function, and let
$x_{m_2} \in F^{m_2}_2$ be uniformly distributed. Then
\begin{equation*}
\begin{aligned}
\sum_{x_{m_2}\in F^{m_2}_2}&\text{cor}^2_{x_{m_1},x_{m_3}}(a\cdot x_{m_1} \oplus c\cdot x_{m_3}\oplus b\cdot f(x_{m_1},x_{m_2},x_{m_3}))\\
 &= 2^{m_1+m_3}\sum_{d\in F^{m_2}_2} \text{cor}^2_{x_{m_1},x_{m_2},x_{m_3}}(a\cdot x_{m_1} \oplus c\cdot x_{m_3} \oplus d\cdot x_{m_2} \oplus b \cdot f(x_{m_1}, x_{m_2}, x_{m_3})),
\end{aligned}
\end{equation*}
for all $a \in F^{m_1}_2$, $c \in F^{m_3}_2$ and $b \in F^{m}_2\setminus \{0\}$, where $m_1, m_2, m_3, m$ are positive integrals.
\end{thm}
\begin{cor}
 Let $f : F^{m_1}_2 \times F^{m_2}_2 \times F^{m_3}_2 \rightarrow F^m_2 $. Then the following are equivalent
\begin{itemize}
 \item [\rm (i)]
   $\text{cor}_{x_{m_1},x_{m_3}}\big((0,...,0,b_{q},0,...,0) \cdot f(x_{m_1},x_{m_2},x_{m_3})\big) = 0 $,  for all  $b_{q} \in F^{q}_2 \setminus \{0\};$
  \item[\rm (ii)] $\text{cor}_{x_{m_1},x_{m_2},x_{m_3}}\big((0,...,0,d_{m_2}, 0,...,0)\cdot x
  \oplus(0,...,0,b_{q},0,...,0) \cdot f(x_{m_1},x_{m_2},x_{m_3})\big) =0$,

 for all $d_{m_2} \in F^{m_2}_2$ and $b_q \in F^q_2\setminus\{0\}.$
\end{itemize}
\end{cor}
Let $a=0\in F_2^{m_1}$, $c=0\in F_2^{m_3}$ and $b=(0,...,0,b_{q},0,...,0)$, the equivalence of (i) and (ii) follows from Theorem 2.2.
\section{\large\bf Some integral distinguishers on 6-round of E2 }

In this section, we show a series of integral distinguishers of 6-round E2.  Firstly, we derive several types of zero-correlation linear approximations for 6-round E2 following the properties on the propagation of linear masks over basic block
cipher operations proposed in [14].

We assert that the 6-round  linear approximations
$$(0,0,0,0,0,0,0,0; 0,0,0,0,0,0,0,b)\rightarrow(0,0,0,0,0,h,0,0; 0,0,0,0,0,0,0,0)$$
have zero-correlation, where $b$ and $h$ denote any non-zero value.

Consider that the input masks$(0,0,0,0,0,0,0,0; 0,0,0,0,0,0,0,b)$ will result that the input mask for
 $L_3$ is $(f_1,f_2,f_3,f_4,f_5,f_6,f_7,f_8\oplus b)$ in the forward direction, where $f_i$, $1 \leq i\leq 8$ denotes a non-zero value. Similarly, in the backward direction, we can get that the output mask of $L_3$ is $(0,g_1,g_2,g_3, 0,0,g_4,g_5)$ from the output $(0,0,0,0,0,h,0,0; 0,0,0,0,0,0,0,0)$ where $g_i$,$1 \leq i\leq 5$ also denotes a non-zero value. Then, we have $f_1=0$ and $f_5=0$ and therefore $e_1=0$ and $e_5=0$. According to the relations between $e_i$ and $d_i$, the following equations
should hold,
$e_1=d_3\oplus d_4 \oplus d_5 \oplus d_6 \oplus d_8 =0 $, $e_5=d_3\oplus d_4 \oplus d_5 \oplus d_8=0 $. It is easy to deduce that $d_6 = 0$, which contradicts that $d_6\neq0$. As a result, the linear hull is  a zero-correlation linear hull, see Figure 2.
\begin{figure}
 \centering
  \includegraphics[width=8cm]{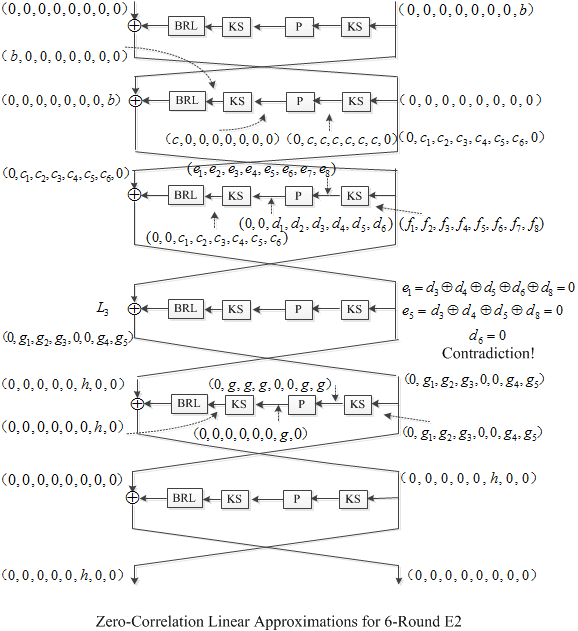}
  \caption{Zero-correlation linear approximations of 6-round E2}
\end{figure}

By Corollary 2.1, we can deduce some  6 round integral distinguishers  from the zero-correlation linear approximations of E2, which are described in the following property.
\begin{prop} Choose a set of $2^{120}$  input of the $r$ round , where the values of
byte $R_r[8]$  are chosen and other bits are chosen to be arbitrary
constants. Encrypt the chosen ${2^{120}}$ values $6$ rounds,  then, each
of the $2^{8}$ possible values of $L_{r+6}[6]$ occurs exactly $2^{112}$ times.
\end{prop}

The integral distinguishers in the above property have much stronger ability to distinguish the right and wrong keys. Let
$ F : F^{120}_2 \rightarrow F_2^8$ be a random vectorial Boolean function and the sets $A_j= \{x_j \in F_2^{120}| F(x_j)=y_j\}$, where $y_j \in F_2^8$, $1 \leq j\leq 2^8$, then the probability of the random vectorial Boolean function satisfying $|A_j|=2^{112}$, for each $1 \leq j\leq 2^8$ is about $2^{-14531}$, which is extremely small.
\section{\large\bf Key-recovery attacks on 6/7-round E2-128/192  with $IT$ and $FT$}

In this section, we present integral attacks on 6-round E2-128 and 7-round E2-192 with $IT$ and $FT$ using
 the integral distinguisher in Property 3.1. We first show a property about the modular multiplication over $F^{32}_2$, which is adopted both in $IT$ and $FT$.

\begin{prop}
Denote 32-bit input, output and subkey of the modular multiplication over $F^{32}_
2$ in $IT$ function as $(x_4, x_3, x_2, x_1)$, $(y_4, y_3, y_2, y_1)$ and $(k_4, k_3, k_2, k_1)$
respectively, where $x_1$, $y_1$ and $k_1$ are the least significant bytes. Then the output
byte $y_i$ is only related to $x_1,.., x_i$ and $k_1,.., k_i$. Moreover, If $x_1$ is fixed as a constant,
and $(x_4,x_3,x_2)$ traversal $F_2^{24}$, then $y_1$ is a corresponding constant, and $y_4,y_3,y_2$ traversal $F_2^{24}$.
\end{prop}

Property 4.1 can be proved easily according to the property of modular multiplication operation in $F^{32}_2$,  which means that
when we want to compute $y_i$, we only need to guess $k_1,...,k_i$ other than guess all 32-bit subkey
values. Meanwhile, there is no need to obtain all 32-bit input value either, the
knowledge of the value $x_1,..., y_i$ is sufficient to compute $y_i$. see [25].
In addition, to construct special structures of the output of $IF$ function that some
bits are arbitrary constants, while the other  bits take all possible values, we only need to
construct  special structures of the input of $IF$ function satisfy corresponding properties.

\subsection{\large\bf Key-recovery attacks on 6-round E2-128  with $IT$ and $FT$}

To attack 6-round E2-128 with $IT$ and $FT$, the 6-round integral distinguishers from Figure 2 start from  round 1 and end
at  round 6. The $IT$ function is added before and $FT$ function is
appended after the integral distinguishers, refer to Figure 3(a).
The partial encryption and decryption using the partial sum technique are
proceeded as follows.

1. Choose a set of $2^{120}$ plaintexts to construct a structure,
where  $P_R[4]$ are chosen to be arbitrary constants over $F_2^8$, while the other 120 bits take all possible values of $F_2^{120}$. Allocate 128-bit counters $V_1[x_1]$ for $2^{24}$ possible values of $x_1= C_L[2,3,4]$ and initialize them to zero. For the corresponding ciphertexts after 6 round encryption with $IT$ and $FT$, extract the value of $x_1$ and increment the corresponding counter $V_1[x_1]$. The time complexity
of this step is $2^{120}$ memory accesses to process the chosen $P C$ pairs. We assume that processing each $P C$ pair is equivalent to 1/4 round encryption, then the time complexity of this step is about $2^{120} \times 1/4 \times 1/6 \approx 2^{115.4} $ 6-round encryptions.

2. Allocate 128-bit counters $V_2[x_2]$ for $2^{8}$ possible values of $x_2 = L_6[2]$ and initialize them to zero.
Guess $K_{10}[2,3,4]$ and $K_9[2,3,4]$ and partially encrypt $x_1$ to get the value of $x_2$, then update the corresponding counter by
$V_2[x_2] + = V_1[x_1]$. The computation in this step is much simpler than 1/2
round encryption and is proceeded about $2^{24}\times 2^{24} \times 2^{24}\times 1/2 = 2^{71}$.

\begin{figure}
 \centering
  \includegraphics[width=10.5cm]{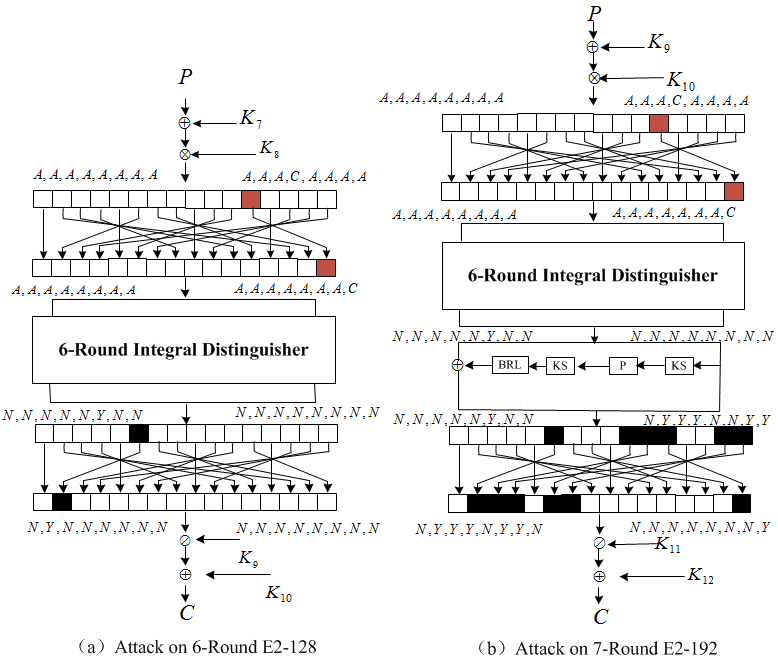}
  \caption{Key-recovery attacks on 6 and 7-round E2-128/192  with $IT$ and $FT$}
\end{figure}

3. After Step 2, 48 key bits have been guessed. If there exists $x_2 \in F_2^8$,  $V_2[x_2] \neq 2^{112}$, discard the guessed keys and guess another subkey  until we get the correct subkey

In this attack,there are 48-bit key value guessed during the encryption phase, and only the right key candidates survive in the wrong key filtration. The dominant complexity of Step 1 is no more than $2^{115.4}$ 6-round  encryptions.  In total, the data complexity is about $2^{120}$ chosen plaintexts, the time complexity is about $2^{115.4}$ 6-round encryptions  and the memory requirement are $2^{28}$ bytes for counters.

\subsection{\bf Key-recovery attack on 7-round E2-192  with $IT$ and $FT$}
Let the 6-round  integral distinguishers cover from round 1 to round 6, and add  $IT$ function before and append one round and $FT$ function after. The details of the attack 7-round E2-192 with $IT$ and $FT$ are illustrated as follows, also see  Figure 3(b).

1. Choose a set of $2^{120}$ plaintexts to construct a structure,
where  $P_R[4]$ are chosen to be arbitrary constants over $F_2^4$, while the other 120 bits take all possible values of $F_2^{120}$. Allocate 128-bit counters $V_1[x_1]$ for $2^{56}$ possible values of $x_1= C_L[2,3,4,6,7,8]\|C_L[8]$ and initialize them to zero. For the corresponding ciphertexts after 7 round encryption with $IT$ and $FT$, extract the value of $x_1$ and increment the corresponding counter $V_1[x_1]$. The time complexity of this step is $2^{120}$ memory accesses to process each chosen $P C$ pairs. We assume that processing each $P C$ pair is equivalent to 1/4 round encryption, then the time complexity of this step is about $2^{120} \times 1/4 \times 1/7 \approx 2^{115.3} $ encryptions.

2. Allocate 128-bit counters $V_2[x_2]$ for $2^{48}$ possible values of $x_2 =R_7[6]\| L_7[2,3,4,7,8]$ and initialize them to zero.
Guess $K_{10}[2,3,4,6,7,8,16]$ and $K_{11}[2,3,4,6,7,8,16]$ and partially encrypt $x_1$ to get the value of $x_2$, then update the corresponding counter by $V_2[x_2] + = V_1[x_1]$. The computation in this step is no more than 1/2
round encryption and is proceeded about $2^{56}\times 2^{56} \times 2^{56}\times 1/2 \times 1/7 = 2^{164.2}$.

3. Allocate 128-bit counters $V_3[x_3]$ for $2^{48}$ possible values of $x_3 =R_7[6]\| L_7[3,4,7,8]\|X[2]$ and initialize them to zero.
Guess $K^{1}_{7}[2]$ and partially encrypt $x_2$ to get the value of $x_3$, then update the corresponding counter by $V_3[x_3] + = V_2[x_2]$. The computation in this step is no more than   1/8
round encryption and the time complexity is about $2^{56}\times 2^{56} \times 2^{8}\times 2^{48}\times 1/8 \times 1/7 = 2^{162.2}$.

The following steps in the partial encryption and decryption phase are similar
to Step 3. Thus, to be consistent, we use Table 2 to show the
details of each partial encryption and decryption step.

\begin{table}[tbp]
\centering
\scriptsize
\begin{tabular}{cccll}
\hline
Step & Guess Keys & Complexity & Computed States  \\
\hline
\vspace{-0.1in}4 &$K^{(1)}_7[3]$ &  $2^{128}\times 2^{48}$& $x_4=R_7[6]\| L_7[4,7,8]\|X[2]\oplus X[3]$ \\
\vspace{-0.1in}5 &$K^{(1)}_7[4]$ &  $2^{136}\times 2^{40}$& $x_5=R_7[6]\| L_7[7,8]\|X[2]\oplus X[3]\oplus X[4]$ \\
\vspace{-0.1in}6 &$K^{(1)}_7[7]$ &  $2^{144}\times 2^{32}$& $x_6=R_7[6]\|L_7[8]\|X[2]\oplus X[3]\oplus X[4]\oplus X[7]$ \\
\vspace{-0.1in}7 &$K^{(1)}_7[8]$ &  $2^{152}\times 2^{24}$ & $x_7=R_7[6]\|\|X[2]\oplus X[3]\oplus X[4]\oplus X[7]\oplus X[8]$ \\
               8 &$K^{(2)}_7[7]$ &  $2^{160}\times 2^{16}$ & $x_8=R_6[6]$ \\
\hline
\end{tabular}

\caption{Partial Encryption and Decryption of the Attack on 7-Round E2-192}
\end{table}

9. After Step 8, 168 key bits have been guessed. If there exists any $x_9 \in F_2^8$, $V_9[x_9] \neq 2^{112}$, then, discard the guessed keys and guess another subkey  until we get the correct subkey

In this attack, there are 168-bit key value guessed during the encryption phase, and only the right key candidates can survive in the wrong key filtration. The dominant complexities of Step 3 to 8 are no more than $2^{172.5}$ 7-round  encryptions totally.  Then, the data complexity is about $2^{120}$ chosen plaintexts, the time complexity is about $2^{172.5}$ 7-round encryptions and the memory requirement are $2^{60}$ bytes for counters.

\section{\large\bf Conclusion }
In this paper, we evaluate the security of E2 block cipher with respect to the technique of integral cryptanalysis.
 We deduce some 6-round  integral distinguishers from new-revealed 6-round zero-correlation linear approximations of E2 by improving the relation between zero correlation linear approximations and integral distinguishers. Besides we give the first
integral attack on the 6 round  E2-128 and the 7 round E2-192 with $IT$ and $FT$ functions taken into consideration. The two attacks need $2^{115.4}$ encryptions with $2^{120}$ chosen plaintexts and $2^{172.5}$ encryptions with $2^{120}$ chosen plaintexts, respectively.

\vspace{0.1in}

\leftline{\bf References} \bigbreak
\def\REF#1{\par\hangindent\parindent\indent\llap{#1\enspace}\ignorespaces}
\footnotesize
\small
\REF{[1]} J. Daemen, L.Knudsen,  V. Rijmen, The block cipher Square. In: Biham, E.(ed.), FSE 1997, LNCS, vol. 1267, Springer, Heidelberg, 1997, pp. 149-165.
\REF{[2]} L. Knudsen,  D. Wagner, Integral cryptanalysis. In: Daemen, J., Rijmen, V.(eds.), FSE 2002, LNCS, vol. 2365, Springer, Heidelberg 2002, pp. 112-127.
\REF{[3]} S. Lucks, Attacking seven rounds of Rijndael under 192-bit and 256-bit keys. In: Proc. 3rd AES Candidate Conf., 2000, pp. 215-229.
\REF{[4]} A. Biryukov, A. Shamir, Structural Cryptanalysis of SASAS. In: Pfitzmann, B.(eds.), EUROCRYPT 2001, LNCS, vol. 2045, Springer, Heidelberg 2001, pp. 394-405.
\REF{[5]} E.Biham,  A. Shamir, Differential cryptanalysis of DES-like cryptosystems. In Menezes, A., Vanstone, S.A.(eds.): CRYPTO 1990. Vol. 537 , LNCS., Springer Heidelberg 1991, pp. 2-21
\REF{[6]}  M. Reza Z'aba, H. Raddum,  M. Henricksen, E.Dawson, Bit-Pattern Based Integral Attack. In: Nyberg, K. (eds.), FSE 2008. LNCS, vol. 5086, Springer, Heidelberg 2008, pp. 363-381.
\REF{[7]}  Y. Li, W. Wu, L. Zhang, L. Zhang, Improved Integral Attack on Rijndael. Journal of Information Science and Engineering (JISE), to appear.
\REF{[8]} Y. Li, W. Wu, L. Zhang,  Integral Attacks on Reduced-Round ARIA Block Cipher. In: Kwak, J., et al. (eds.) ISPEC 2010. LNCS, vol. 6047, Springer, Heidelberg 2010, pp. 19-29.
\REF{[9]}   X. Sun, L. Lai, Improved Integral Attacks on MISTY1. In: S.A., (eds.), Volume 5867, LNCS, 2009, pp. 266-280.
\REF{[10]} J. Daemen,  V. Rijmen,  AES proposal: Rijndael. In AES Round 1 Technical Evaluation CD-1: Documentation. NIST, August 1998. http://www.nist.gov/aes.
\REF{[11]} M. Matsui,: Linear cryptanalysis method for DES cipher. EUROCRYPT'93, Vol. 765, LNCS, Springer 1993,  pp. 386-397.
\REF{[12]} A.Biryukov, C.D. Canniere, M. Quisquater, On Multiple Linear Approximations. In: Franklin, CRYPTO 2004, LNCS, volume. 3152, Springer 2004, pp. 1-22.
\REF{[13]} M. Hermelin, J.Y. Cho,  K. Nyberg, Multidimensional Extension of Matsui's Algorithm 2, FSE 2009, LNCS, volume. 5665, Springer 2009, pp. 209-227.
\REF{[14]}  A. Bogdanov, V. Rijmen, Linear Hulls with Correlation Zero and Linear Cryptanalysis of Block Ciphers. Designs, Codes and Cryptography March 2014, Volume 70, Issue 3, pp. 369-383.
\REF{[15]}  E. Biham , A. Biryukov  A. Shamir , Cryptanalysis of Skipjack Reduced to 31 Rounds using Impossible Differentials,  EUROCRYPT 1999, LNCS, volume 1592, Springer, pp. 12-23.
\REF{[16]}  F. Chabaud, S.Vaudenay,  Links Between Differential and Linear Cryptanalysis. In De Santis, A. ed: EUROCRYPT-94, Vol. 950, LNCS, Springer 1994, pp.356-365.
\REF{[17]} C, Blondeau, K. Nyberg, New Links between Differential and Linear Cryptanalysis. Lecture Notes in Computer Science, EUROCRYPT 2013, Volume 7881, 2013, pp. 388-404.
%\REF{[17]} E. Biham, A. Biryukov, and A. Shamir, Cryptanalysis of Skipjack Reduced to 31 Rounds using Impossible Differentials,  EUROCRYPT¡¯99, %LNCS 1592, pp. 12-23, Springer-Verlag.
%\REF{[17]} Leander, G.: On Linear Hulls, Statistical Saturation Attacks, PRESENT an Cryptanalysis of PUFFIN. In Paterson, K.G. ed.: EUROCRYPT 2011. Vol. 66 of LNCS, Springer (2011) 303-322
%\REF{[19]} Collard, B., Standaert, F.X.: A statistical saturation attack against the block cipher PRESENT. In Fischlin, M. ed.: CT-RSA 2009. Vol. 5473 of LNCS, Springer (2009) 195-210
%\REF{[20]} Hermelin, M., Cho J. Y., and Nyberg, K.: Multidimensional extension of Matsui¡¯s
%algorithm 2. In Dunkelman, O. ed.: FSE 2009, Vol. 5665 of LNCS, Springer (2009)209-227
\REF{[18]} A. Bogdanov, G. Leander,  K. Nyberg, M. Wang, Integral and Multidimensional Linear Distinguishers with Correlation Zero. In Wang, K.,Sako, K., (eds), ASIACRYPT 2012 Vol. 7658, LNCS., Springer 2012, pp. 244-261.
%\REF{[24]} Nyberg, K.: Linear approximation of block ciphers. In De Santis, A. ed: EUROCRYPT¡¯94, Vol. 950 of LNCS, Springer (1995) 439¨C444.
%\REF{[25]} Bogdanov, A.,  Rijmen, V.: Linear Hulls with Correlation Zero and Linear Cryptanalysis of Block Ciphers. Designs, Codes and Cryptography, Springer, US, 2012, pp.1-15.
\REF{[19]} M. Kanda,  S. Moriai,  K. Aoki,   H. Ueda, Y. Takashima, K. Ohta, T. Matsumoto,
E2-a new 128-bit block cipher. IEICE Transactions Fundamentals of Electronics,
Communications and Computer Sciences 2000, E83-A(1), pp. 48-59.
\REF{[20]} M. Matsui,    T. Tokita,  Cryptanalysis of a Reduced Version of the Block Cipher E2.
FSE 1999, LNCS, vol. 1636, Sringer 1999, pp. 56-71.
\REF{[21]}  S. Moriai,   M. Sugita,   K. Aoki,  M.Kanda, Security of E2 against Truncated Differential Cryptanalysis. SAC 1999, LNCS, vol. 1758, Springer 2000, pp. 106-117.
\REF{[22]} K. Aoki,  M. Kanda,  Search for Impossible Differential of E2, 1999.
(Available from: http://csrc.nist.\\
gov/encryption/aes/round1/comment).
\REF{[23]} Y. Wei,  P. Li,   B. Sun, C. Li,  Impossible Differential Cryptanalysis on Feistel Ciphers with SP and SPS Round Functions. ACNS 2010, LNCS, vol. 6123, Springer 2010, pp. 105-122.
\REF{[24]}  Y. Wei, X.Yang, C. Li, W. Du,  Impossible Differential Cryptanalysis on Tweaked E2. Concurrency and Computation: Practice and Experience, 2013.
\REF{[25]}  L. Wen, M. Wang,   A. Bogdanov,  Multidimensional Zero-Correlation Linear Cryptanalysis of E2. to appear.
\REF{[26]} K. Nyberg,  Linear approximation of block ciphers. In De Santis, A. ed, EUROCRYPT¡¯94, Vol. 950,  LNCS, Springer 1995, pp.439-444.

\end{document}